\documentclass[reprint,
 amsmath,amssymb,
 aps,
prl,
]{revtex4-2}

\usepackage{graphicx}  
\usepackage{dcolumn}   
\usepackage{bm}        
\usepackage{epsfig,amsmath,amssymb,verbatim,mathrsfs,array,layout,textcomp,amssymb,latexsym,slashed}
\usepackage{xcolor}
\usepackage{comment}
\usepackage{siunitx}

\usepackage[colorlinks=true,citecolor=blue,urlcolor=blue,linktocpage=true,
linkcolor=blue]{hyperref}
\usepackage[utf8]{inputenc}
\usepackage{multirow}
\usepackage[capitalize]{cleveref}

\usepackage{comment}

\usepackage{physics} 

\usepackage{tikz}

\usepackage{placeins}

\DeclareMathOperator{\sinc}{sinc}

\begin{document}

\title{%
  \texorpdfstring
    {Is the Conventional Picture of Coherence Time Complete?\\
     Dark Matter Recoherence}
    {Is the Conventional Picture of Coherence Time Complete? Dark Matter Recoherence}%
}

\author{Chaitanya Paranjape}%
\email{chaitanya.paranjape@weizmann.ac.il}

\author{Gilad Perez}
\email{gilad.perez@weizmann.ac.il}

\author{Wolfram Ratzinger}%
\email{wolfram.ratzinger@weizmann.ac.il}

\author{Somasundaram Sankaranarayanan}%
\email{somasundaram.sankaranarayanan@weizmann.ac.il}

\affiliation{%
 Department of Particle Physics and Astrophysics,\\
 Weizmann Institute of Science, Rehovot, Israel 7610001
}%

\date{\today}

\begin{abstract}
    The local Solar gravitational potential forms a basin for ultralight dark matter (ULDM) with discrete energy levels. Even if barely populated, it introduces a new characteristic timescale in DM dynamics. 
    This necessitates a generalization of the notion of coherence time. We find that, at long times, the phenomenon of ``recoherence" emerges, whereby a subcomponent of ULDM exhibits a formally divergent coherence time. The fact that this generalized coherence time can significantly exceed the naive estimate implies an enhanced sensitivity for dark matter searches that accumulate data over extended observation periods.
\end{abstract}

\maketitle
\section{\label{sec:Introduction}Introduction}
While there exist a large number of astrophysical and cosmological observations of dark matter (DM) ~\cite{Bertone:2016nfn,Weber_2010,Nesti:2012zp,Bovy:2012tw,Read:2014qva,Evans:2018bqy,Necib:2018igl}, its nature remains one of the largest mysteries in physics. Models of ultralight DM (ULDM)
provide arguably the simplest explanation for the origin of DM. Due to its astronomical occupation and weak coupling, ULDM behaves as a classical field, which oscillates with a frequency equal to the particle's total energy \cite{Bao:2025nsd}. Being non-relativistic, the ULDM frequency is close to its mass with small corrections due to kinetic and/or potential energy. 
Therefore, the signal induced by ULDM in direct experimental searches is to a good approximation, harmonic and deterministic, setting it apart from noise and therefore enhancing the sensitivity.
This enhancement is, however, limited by the stochastic nature of the DM dynamics~\cite{Foster:2017hbq,Centers:2019dyn,Lisanti:2021vij,Kim:2021yyo,Flambaum:2023bnw,Kim:2023kyy,Kim:2023pvt,Cheong:2024ose,Gan:2025icr}\,, associated with the fact that it is known to have a finite coherence time, after which its phase and amplitude information is lost. 
The standard lore is that the ULDM coherence time is given by ${\cal O}\left(1/m\sigma^2\right)$, with $m$ being the DM mass and $\sigma\sim 10^{-3}$ being the dispersion of the galactic DM velocity distribution. 
This  lore is missing a crucial piece of available information, namely that, locally, the ULDM is subject to the gravitational potential of the Sun. It implies that in addition to the continuous set of scattering states, there is a set of discrete bound states that ULDM may populate~\cite{Banerjee:2019epw,Banerjee:2019xuy}. In~\cite{ Kaup:1968zz,Ruffini:1969qy,1991ApJ...368..610G,Kolb:1993zz,Xu:2008ep,Peter:2009mi,Peter:2009mm,Lasenby:2020goo,Lundberg:2004dn,VanTilburg:2020jvl,Budker:2023sex}\,,
several gravitational and non-gravitational dynamical mechanisms where discussed in which these discrete states are populated. 

As we show below, the presence of these discrete states may dramatically change how we think about the ULDM coherence time and more importantly affect the expected sensitivity of current and near future ULDM experiments. 
We show that even a small amount of DM bounded to spatially confined systems leads to a formal divergence of the commonly considered coherence time.
We therefore introduce a straightforward generalization of coherence time that takes into account the time an experiment is run for. 
While being finite, it may still vastly exceed the common coherence time resulting in significantly improved experimental sensitivities.

We obtain that, in the limit of long observation time, the generalized coherence time starts growing with the observation time, once the energy levels get resolved. 
We coin this phenomenon DM-\textit{recoherence}. 

\begin{figure}[h!]
    \includegraphics[scale=0.48]{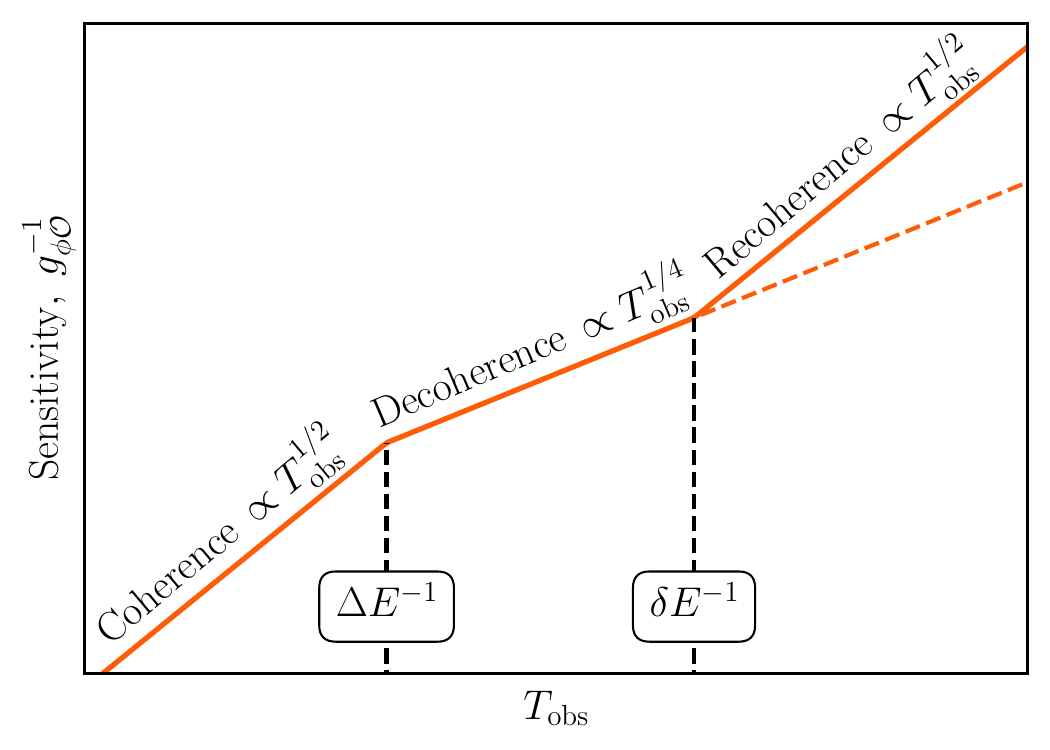}
     \caption{Evolution of the sensitivity to the effective DM coupling in a broadband search as a function of observation time. The energy spread $\Delta E$ of states populated by DM leads to decoherence and a deterioration of the sensitivity. Bound states with finite energy spacing $\delta E$ eventually lead to recoherence and the same scaling as in the coherent regime is recovered.}
\label{fig:Coherence_Revelation_Sensitivity_Scaling_Schematic}
\end{figure} 

In \cref{fig:Coherence_Revelation_Sensitivity_Scaling_Schematic} we show the schematic scaling of the sensitivity to the effective coupling of a scalar $\phi$ constituting DM to the standard model in a broadband search. Due to the DM populating multiple levels with energies spread by $\Delta E$\,, the DM signal decoheres after an observation time, $T_\mathrm{obs}\sim \Delta E^{-1}$\,. Beyond this point the sensitivity scales as $T_\mathrm{obs}^{1/4}$ rather than $T_\mathrm{obs}^{1/2}$\,. DM in discrete bound states, with a typical energy spacing $\sim \delta E$\,, eventually leads to the phenomenon of recoherence, beyond which the scaling $T_\mathrm{obs}^{1/2}$ is recovered.
To give the readers a rough idea regarding the relevant scales, consider ULDM with a mass of $10^{-14}\,$eV populating the Sun's gravitational potential at $5\%$ of background galactic DM density, then the DM signal decoheres after about $\sim 1/m \sigma^2 \sim$ a day, and recoheres after approximately $\sim 400/m \sigma^2 \sim$ 1 year.
This implies that small fractions of DM in bound states may drive discovery prospects in ongoing experiments that are run over long periods, as we show below.

\section{\label{sec:Generalized coherence time}Generalized coherence time}

A real scalar field $\phi$ in the non-relativistic regime, where 
$\dot\phi \sim m\phi \gg \nabla\phi$, can be written as 
$\phi = (\psi e^{-imt} + \mathrm{h.c.})/\sqrt{2m}$,
with the slowly varying field $\psi$ obeying the Schr\"odinger equation, $i\dot\psi = \left[-\nabla^2/(2m) + V(\vec r)\right]\psi$\,.
The effective potential $V(\vec r)=m\Phi(\vec r)+\delta m(\vec r)$ encodes both the gravitational potential and possible interactions, see for instance \cite{Budker:2023sex,Gorghetto:2024vnp,Banerjee:2025dlo,delCastillo:2025rbr}.

To make the discussion simpler, we assume negligible self-interactions and ignore the field’s self-gravity, so the Schr\"odinger equation is linear and the field may be expanded into eigenmodes $\psi_i(\vec r)$ with energies $\omega_i = m + E_i$. The classical field may then be written as,
$\phi = \sum_i (\alpha_i \psi_i e^{-i\omega_i t} + \mathrm{h.c.})/\sqrt{2m}$,
where the complex amplitudes $\alpha_i$ are Gaussian random variables with variances fixed by the occupation numbers $N_i$ \cite{Foster:2017hbq,Kim:2021yyo,Kim:2023pvt,Cheong:2024ose}. Within this stochastic framework, the primary observables are the power spectrum and the correlation functions of the field. Our focus here is the coherence time.

The coherence time represents the interval over which the field retains significant self-correlation. In this work we adopt the definition,
\begin{equation}
    \tau_{\rm coh}(T_{\rm obs})
    = \int_{-T_{\rm obs}}^{T_{\rm obs}} d\tau\,
      \left[\frac{\Gamma(\tau)}{\Gamma(0)}\right]^2 ,
    \label{eq:Coh_time_NEW_definition}
\end{equation}
where $\Gamma(\tau)=\langle \phi(0)\phi(\tau)\rangle$ is the fields auto-correlation function. This coincides with e.g. the standard definition in quantum optics in the limit, $T_\mathrm{obs}\rightarrow \infty$ \cite{Mandel:1995seg,Cheong:2024ose}.
This standard definition
is sufficient to capture the relevant dynamical scales of the problem when the spectrum is continuous and is characterized by a single scale $\sim \Delta E\,$. However, in systems where the discrete level spacing $\sim \delta E\,$ becomes resolvable on observational time scales, taking the integration limits to be $\pm \infty$ fails to capture the existence of distinct coherence regimes. 

To discuss how the different coherence regimes originate from the modified definition in Eq.~\eqref{eq:Coh_time_NEW_definition}, we need the auto-correlation of the field. Making use of the expansion into mode functions, the auto-correlation can be evaluated as,
\begin{equation}
    \Gamma(\tau) = \sum_{i} \frac{\abs{\psi_i}^2}{m} N_i \cos(\omega_i\tau) \,, \label{eq:Auto-correlator} 
\end{equation} where the sum runs over all modes $i$. The coherence time is then given as, 
\begin{equation}
    \frac{\tau_{\mathrm{coh}} (T_{\text{obs}})}{T_{\mathrm{obs}}} = \frac{\sum_{i} \abs{\psi_i}^2 N_i \sum_j \abs{\psi_j}^2  N_j \sinc( (E_i - E_j) T_{\mathrm{obs}})  }{\left( \sum_{i} \abs{\psi_i}^2 N_i  \right)^2} \,, \label{eq:Coherence_fraction_discrete} 
\end{equation} 
where we have assumed observation times long enough to sample at least a single DM oscillation, $m T_{\mathrm{obs}} \gtrsim 2\pi$. The $\sinc$-function is defined as, $\sinc(x) = \sin(x)/x$. Let us now analyze the behavior of the coherence time in different  regimes of the observation time. 

\subsection*{%
  \texorpdfstring
    {Coherence: $T_\mathrm{obs} \lesssim \Delta E^{-1}$}
    {Coherence: Tobs <= 1/Delta E}%
}
In the non-relativistic regime the energies of the different modes $E_i$, span some range, $\Delta E\ll m$. For observation times shorter than the inverse of this spread, the $\sinc$-function in \cref{eq:Coherence_fraction_discrete} can be approximated as $\sinc\approx 1$. This results in our generalized coherence time coinciding with the observation time, $\tau(T_\mathrm{obs})=T_\mathrm{obs}$\,. To this end the dark matter signal in the detector can be approximated as a single cosine, $\phi\propto \cos(mt)$\,.

\subsection*{%
  \texorpdfstring
    {Decoherence: $\Delta E^{-1} \lesssim T_\mathrm{obs} \lesssim \delta E^{-1}$}
    {Decoherence: 1/Delta E <= Tobs <= 1/delta E}%
}
Once the observation time exceeds the inverse of the spread in energies, the time evolution of the field differs from a single cosine and we have to deal with the full correlation function in \cref{eq:Auto-correlator} to relate measurements separated by these time-scales. Generically the correlation is poised to decrease after these times, since the different modes contributing to the measurement acquire $O(1)$ phase differences that lead to deconstructive interference. In the simplest case the power spectrum of DM consists of a single peak with width $\Delta E$\,. 
In this case the generalized coherence time saturates at, $\tau(T_\mathrm{obs})\approx\Delta E^{-1}$\,, in agreement with the common lore. This picture holds true e.g. for ULDM in the galactic halo, where $\Delta E\simeq m\sigma^2$ with $\sigma$ denoting the DM velocity dispersion \cite{Foster:2017hbq}.

This simple picture needs to be modified if for example a significant fraction of DM is in a cold stream with a smaller velocity dispersion, in which case two different coherence time scales would be of relevance. In case the spectrum is not continuous, even more drastic effects appear, once the observation time becomes long enough that the different energy levels separated by $\sim\delta E$ can be resolved.

\subsection*{%
  \texorpdfstring
    {Recoherence: $\delta E^{-1} \lesssim T_\mathrm{obs}$}
    {Recoherence: 1/delta E <= Tobs}%
}
Once the various energy levels can be resolved, $T_\mathrm{obs} \gtrsim \delta E^{-1} $\,, the sinc-function can be approximated as $\sinc( (E_i - E_j) T_{\mathrm{obs}})\sim\delta_{E_i,E_j}$\,. In this case the right-hand side of \cref{eq:Coherence_fraction_discrete} becomes constant again, which results in the coherence time growing proportional to the observation time and therefore to a divergence when considering $T_\mathrm{obs}\rightarrow\infty$. The rate of growth is however suppressed with respect to the coherent case by the effective number of populated non-degenerate energy levels $N_\mathrm{nd}$ contributing to the sum in \cref{eq:Coherence_fraction_discrete}, $\tau(T_\mathrm{obs})\approx T_\mathrm{obs} /N_\mathrm{nd}$\,. The effective number of these non-degenerate levels is related to the typical spacing between the levels $\delta E$ via $N_\mathrm{nd}\approx \Delta E/\delta E$\,.

For the galactic halo, the transition time $\delta E^{-1}$ is of the order of the Sun’s orbital period around the galaxy, far exceeding the reach of any realistic experiment; consequently, the standard treatment remains valid, and the discussion of recoherence becomes irrelevant. 
In contrast, for a Solar or a terrestrial halo, the discrete structure becomes relevant on observable timescales, making the consideration of recoherence essential. To illustrate the emergence of recoherence precisely, we present three concrete examples. Below we examine a finite three-dimensional box with discrete momenta as a toy model. Subsequently, we analyze physical scenarios involving the bound states of a Solar DM halo in the background of the galactic halo, where we expect recoherence to improve sensitivities.

\section{\label{sec:Generalized coherence examples}Generalized coherence examples}

\subsection{\label{sec:Free quantum gas in a 3D box}Free quantum gas in a 3D box}

Let us first consider the toy example of a free, non-relativistic quantum gas inside a 3D cubic box of volume $L^3$. We impose periodic boundary conditions such that the momenta are quantized, $\vec{k} = (2 \pi / L) \vec{n}\,, n_i \in \mathrm{Z}$, with wavefunctions given by simple plane waves $\psi_{\vec{k}}(\vec{x}) = L^{-3/2} e^{i \vec{k} \cdot \vec{x}}$, and corresponding energies $E_{\vec{k}} = \frac{\vec{k}^2}{2m}$. Assuming a Maxwell-Boltzmann distribution for the population $N_{\vec k}\propto\exp(-\beta E_{\vec k})$ with effective temperature \,$\beta^{-1}$\,, one can write down the auto-correlation of the field as described in Eq.~\eqref{eq:Auto-correlator}, 
\begin{equation}
    \Gamma(\tau) \propto \sum_{\vec{n}} \frac{e^{-\beta E_{\vec{n}}}}{m L^3} \cos((m + E_{\vec{n}}) \tau) \,. 
\end{equation}  
In this particular example, the typical spread in energies is controlled by the temperature, $\Delta E \sim \beta^{-1}$, and the typical energy spacings depend upon the box size as all energy levels are multiples of, $\delta E = 2 \pi^2 / (m L^2)$. Thus we expect decoherence to occur at time $T_\text{obs} \gtrsim \beta$ and recoherence at $T_\text{obs} \gtrsim \delta E^{-1}$\,.

We can further consider a cuboid box with irrational ratios for side lengths, which removes the degeneracies and leads to the number of non-degenerate levels scaling as $\propto a_L^{3}$, where $a_L$ is a common factor dictating all side lengths. The typical energy spacing therefore scales as, $\delta E\propto a_L^{-3}$ instead of $\propto a_L^{-2}$ as in the degenerate case. This leads to a delayed onset of recoherence as we show in more detail in the supplemental material.

\subsection{\label{sec:Ground state halo}Ground state halo}
Consider the case of DM bound gravitationally to a mass e.g. the Sun, the gravitational atom \cite{Banerjee:2019xuy}. 
The potential is given by  $V(r) = -\alpha/r\,$, with $\alpha = G_N M m$ denoting the gravitational fine structure constant, where $G_N$ is Newton's gravitational constant and $M$ is the attracting mass.
We first consider the case in which only the ground state ($1s$ state) is populated, along with the unbound levels corresponding to dark matter in the galactic halo. Such a system may be formed around the Sun due to DM self-interactions via the mechanism discussed in \cite{Budker:2023sex}. In this mechanism, it was shown that indeed DM in excited bound states tends to relax to the ground state. 

The exact modeling of the auto-correlator in this case is difficult, since the unbound DM populates the continuous part of the spectrum existing of plane waves that get distorted around the central body. This leads to effects like gravitational focusing \cite{Kim:2021yyo}, that result in $\mathcal{O}(1\%)$ variations in the density. For simplicity, we neglect these effects and instead use the autocorrelator for dark matter in the standard halo model, taking the isotropic limit in the absence of such distortions \cite{Cheong:2024ose},
\begin{equation}
    \Gamma_\mathrm{gal} (\tau) =  \frac{\rho^\mathrm{gal}}{m^2} \frac{\cos(m \tau + \frac{3}{2} \tan^{-1}(\tau m \sigma^2))}{(1+ m^2 \sigma^4 \tau^2)^{3/4}} \,,
\end{equation} 
where $\rho^\mathrm{gal}$ is the average galactic energy density of DM and $\sigma$ its velocity dispersion. Including DM in the 1s state the total autocorrelator becomes,
\begin{equation}
    \Gamma (\tau) = \frac{\rho_{1s}}{m^2}\cos\left((m+E_{1s})\tau\right) +\Gamma_\mathrm{gal} (\tau)\,,
\end{equation}
where $\rho_{1s}$ is the DM density in the 1s ground state at the position of the experiment and $E_{1s}=-\alpha^2m/2$ its energy. 
 
If the density in the halo dominates $\rho_{1s}\gg\rho^\mathrm{gal}$\,, the autocorrelation does not decay due to the $1s$ state being perfectly coherent. This results in $\tau_\mathrm{coh}(T_\mathrm{obs})=T_\mathrm{obs}$\,. One may wonder to which degree the coherence of the $1s$ state is an approximation. Since the state is perfectly coherent in the non-interacting theory, its decoherence must be related to perturbative interactions redistributing scalar particles between the levels. It therefore seems likely that the inherent coherence time of a state corresponds to its time of formation or equilibration with other levels. 

Details on possible sources of incoherence can be found in the supplemental material. In particular, we consider the perturbations coming from Jupiters gravitational potential. 
We find that they reduce the ground state lifetime below the age of the Solar system for masses 
$ 3 \cdot\,10^{-15}\,\mathrm{eV} \lesssim m \lesssim 10^{-14}\,\mathrm{eV}$\,, though they remain negligible for the discussion of recoherence.

If the density in the galactic halo dominates on the other hand $\rho_{1s}\ll\rho^\mathrm{gal}$, the generalized coherence time levels off after $T_\mathrm{obs}\gtrsim 1/(m\sigma^2)$\,. However, because the $1s$ state remains individually perfectly coherent, its own contribution to the generalized coherence time continues to grow linearly with $T_\mathrm{obs}$. Therefore, even if the $1s$ Solar halo's density is much smaller than the galactic one, after a time $T_\mathrm{obs}\gtrsim (\rho^\mathrm{gal}/\rho_{1s})^2/(m\sigma^2)$ the signal enters recoherence due to the coherent nature of the $1s$ state and the generalized coherence time starts to grow linearly again. In Fig.~\ref{fig:Solar halo dominating detection prospects} (left), we highlight the region of parameter space where the 1s halo dominates the sensitivity.

\subsection{\label{sec:Virialized halo}Virialized halo}

We next consider the case where all the bound state levels are populated to form a virialized (or thermalized) halo, in a background of galactic halo DM occupying the unbound levels. Trivially, a dominant $1s$ state was always perfectly coherent, however, a virialized halo possesses its own non-trivial coherence behavior, since it is composed of multiple discrete levels. For simplicity let us first neglect the galactic halo contribution such that we can write the auto-correlator as, 
\begin{equation}
    \Gamma_{\text{virial}}(\tau) = \sum_{n l m_l} \frac{\abs{\psi_{nlm_l}(\vec{r})}^2}{m} N_{nlm_l} \cos((m + E_{n l m_l}) \tau) \,,  \label{eq:Virial_Halo_Pure_Correlator}
\end{equation} 
where $\psi_{nlm}(\vec{r})$ are the hydrogen atom wavefunctions labeled by quantum numbers $n,l,m_l$; with $E_{nlm_l}= - m \alpha^2 / 2 n^2$ as the eigenenergies.
We here assume a Maxwell-Boltzmann distribution $N_{nlm_l} \propto e^{-\beta E_{nlm_l}} $, with a high effective temperature $\beta^{-1}\gg mv^2$, where $v = \sqrt{G_N M_\odot / 1\,\mathrm{AU}}$ is the virial velocity at the position of Earth, in the Sun's gravitational potential\,. This is e.g. the case if the bound states are in energetic equilibrium with the galactic halo $\beta^{-1} = m \sigma^2$, since $\sigma\gg v$. We shall estimate the coherence time evolution of this system at a static point at the average distance of the Earth from the Sun, $r = 1\,\text{AU}$. 
The sum in Eq.~\eqref{eq:Virial_Halo_Pure_Correlator}, is dominated by states of a primary quantum number $n_\mathrm{max} \sim \sqrt{r/a_0}$\,, which we also expect to be the typical number of non-degenerate states contributing significantly to the spectrum, $N_\mathrm{nd}\sim n_\mathrm{max}$\,. 
Therefore, we can estimate the expected spread in energies as $ \Delta E \sim m \alpha^2 / n^2_\text{max} = m v^2$\,. We can also estimate the typical energy spacings, $\delta E \sim  \Delta E / N_\mathrm{nd} = m v^2 / \sqrt{r/a_0} = (2 \pi / \text{1 year})$\,. Concluding, we mark the decoherence time scale to be of order $\sim \Delta E^{-1} \sim 1/ m v^2$\,, and the recoherence time scale to be of order $\sim \delta E^{-1} \sim (\text{1 year}/2 \pi)$\,. 

This picture changes slightly when we account for Earth's rotation around the Sun, which lifts the degeneracy in the magnetic quantum number $m_l$. 
Similar to the non-degenerate box, the typical number of non-degenerate states contributing to the spectrum is increased as $N_\mathrm{nd} \sim n^2_\mathrm{max}$. Thus, the onset of recoherence is delayed to $\sim \delta E^{-1} \sim  (\text{1 year}/2 \pi)\, \sqrt{r/a_0}$\,. 
Details and numerical evaluations of the coherence time confirming our analytic estimates for both the ground state and virialized halo can be found in the supplemental material.

Let us finally comment on the case of a sub-dominant virialized halo $\rho^\text{virial} \lesssim \rho^{\text{gal}}$.
For a halo with a large Bohr radius, $r/a_0 \lesssim 1$\,, the halo density will be dominated by the $1s$ state, and thus, the analysis becomes similar to the previous example. However, for smaller Bohr radii, $r/a_0 \gtrsim 1$\,, many states contribute at Earth, and depending upon the halo density and the time of observation, multiple stages of de- and recoherence can be observed. Firstly, since the virial velocity $v$ is much smaller than the galactic velocity, the galactic DM population decoheres followed by recoherence from the virial halo as a whole, since $\Delta E^{-1}\simeq mv^2\ll m\sigma^2$\,. 
This intermediate period of coherence leads to the constant density fraction at large masses for which the Solar halo dominates detection prospects. This fraction is given by the ratio of velocities $v/\sigma\approx0.1$ and can be seen in \cref{fig:Solar halo dominating detection prospects} on the right.
Secondly, the virial halo decoheres followed by recoherence caused by the discrete nature of its energy levels which is resolvable for masses $m\lesssim 3\cdot 10^{-13}$\,eV after 10 years.

\section{\label{sec: implications for experiments} Implications for Experiments}
\begin{figure*}[t]
\centering{\large{Solar halo dominating detection prospects}}
    \includegraphics{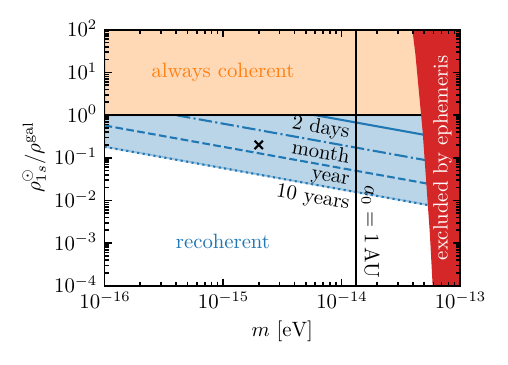}%
    \includegraphics{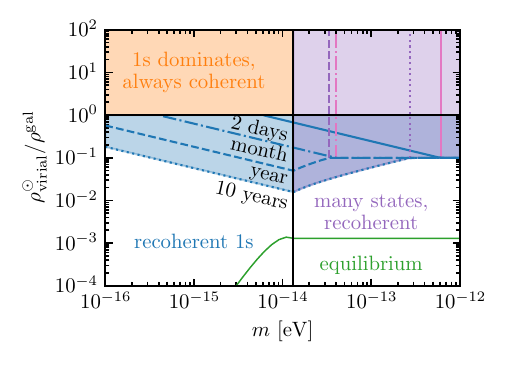}
    \vspace{-0.5cm}
    \caption{In the shaded regions, the Solar halo dominates the detection prospects. This may even be the case if the halo is less dense than the galactic DM at the position of Earth ($\rho^\odot<\rho^\mathrm{gal}$, region below horizontal black line). The vertical black line marks the DM mass $m$ for which the gravitational atom's Bohr radius equals 1\,AU.
    \textbf{Left:} Halo populated only by the 1s ground state. The DM field is always coherent if the Solar halo dominates (orange). If subdominant, it recoheres on timescales longer than the galactic halo coherence time $\sim 1/(m\sigma^2)$ such that even a subdominant halo can enhance the experimental significance for sufficiently long integration times (blue). Extrapolating the halo profile inward excludes the red region from Mercury's ephemeris~\cite{Pitjev:2013sfa}.
    \textbf{Right:} Virialized halo with all gravitational atom states populated. For Bohr radii larger than 1\,AU (left of the black line), the 1s state dominates and the behavior matches the left panel. For smaller Bohr radii, multiple states contribute at Earth, causing recoherence even when the galactic halo is negligible (light purple), and potentially multiple stages of de- and recoherence when it dominates (dark purple). The pink, vertical lines mark the mass at which for small observation times (2 days, month) the Solar halo as a whole decoheres due to its finite width. The purple lines conversely give the mass where the halo recoheres after longer times (year, 10 years) due to the discrete energies. The green line shows the expected Solar halo density if equilibrated with the galactic halo.
    }

    \label{fig:Solar halo dominating detection prospects}
\end{figure*}

The possible increase of the coherence time during a period of recoherence implies an enhanced sensitivity of experiments searching direct interactions of the DM field with the standard model. For concreteness we limit the discussion here to clock comparison searches for linear interactions of a scalar \cite{Uzan:2010pm,Arvanitaki:2014faa,Stadnik:2015kia,Hees:2016gop,Kennedy:2020bac,Campbell:2021mnu,Oswald:2021vtc,Kobayashi:2022vsf,Zhang:2022ewz,Sherrill:2023zah,Banerjee:2023bjc,Fuchs:2024xvc}. We expect however that similar enhancements will appear in other experimental platforms targeted at pseudo-scalars (axions, see~\cite{Kim:2008hd, Graham:2015ouw} and references therein)  as well as searches for quadratic interactions \cite{Banerjee:2022sqg}.

Schematically, the interactions we are interested in are of the form $\, \mathcal{L}\supset g_{\phi\mathcal{O}}\,\phi\,\mathcal{O}_\mathrm{SM}$\,, where $g_{\phi\mathcal{O}}$ is the effective DM coupling suppressing the interaction and $\mathcal{O}_\mathrm{SM}$ is a standard model operator. The operator might be given by the square of a field strength $\mathcal{O}_\mathrm{SM}=FF\,,\,GG$\,, effectively resulting in the electric or strong coupling constant varying with the field, or $\mathcal{O}_\mathrm{SM}=m_f \bar{f}f$ varying the mass of a ferminon $f$\,. Since atomic transitions have different dependency on the coupling constants and masses, this results in a variation of their frequency ratio $\delta\log(\nu_1/\nu_2)=K_{\mathcal{O}}\, g_{\phi\mathcal{O}}\,\phi$\,. Here $K_{\mathcal{O}}$ denotes the relative sensitivity factor to the operator $\mathcal{O}_\mathrm{SM}$ of the two transitions $\nu_{1/2}$ under consideration. 

For a clock comparison test the ratio $\nu_1/\nu_2$ is recorded every time $\Delta t$ for some period $T_\mathrm{obs}$\,. Subsequently, one may Fourier transform the data and search for dark matter in the frequency range $2\pi/T_\mathrm{obs}\lesssim m\lesssim \pi/\Delta t$\,. The measurement errors of subsequent measurements in these experiments are typically uncorrelated. This leads to a white noise power spectrum $P_\mathrm{ns}$, which is independent of the frequency. The goal of a dark matter search is to pick out the frequency modulation caused by the coupling of the scalar out of this noise. We show in the supplemental material that the signal to noise ratio (SNR) of such a search is given by
\begin{equation}
    \mathrm{SNR}^2=\frac{T_\mathrm{obs}\tau_\mathrm{coh}(T_\mathrm{obs})}{2}\frac{\Gamma^2_\mathrm{sig}(0)}{P^2_\mathrm{ns}}\,,
\end{equation}
where $\tau_\mathrm{coh}(T_{\text{obs}})$ is the generalized coherence time introduced in \cref{eq:Coh_time_NEW_definition}. The strength of the signal is characterized above by $\Gamma_\mathrm{sig}(0)=\langle(\delta\log(\nu_1/\nu_2))^2\rangle=K^2_{\mathcal{O}}\,g^2_{\phi\mathcal{O}}\, \langle\phi^2\rangle$\,. The field strength can be related to the local dark matter density $\langle\phi^2\rangle\approx\rho_\mathrm{DM}/m^2$\,. In this way we can express the sensitivity to the effective coupling $g_{\phi\mathcal{O}}$ that can be probed by a given experiment as,
\begin{equation}
    g^{}_{\phi\mathcal{O}}=\left(\frac{T_\mathrm{obs}\tau_\mathrm{coh}(T_\mathrm{obs})}{2}\right)^{-1/4} \left(\frac{\rho_\mathrm{DM}}{P_\mathrm{ns}\mathrm{SNR}_\mathrm{th}}\right)^{-1/2}\frac{m}{K_{\mathcal{O}}}\,, \label{eq:Sensitivity_scaling_equation}
\end{equation}
where $\mathrm{SNR}_\mathrm{th}$ denotes the threshold SNR required for detection or exclusion. We therefore find that the sensitivity grows as $\propto T_\mathrm{obs}^{1/2}$ while the coherence time grows linear with $T_\mathrm{obs}$\,, and $\propto T_\mathrm{obs}^{1/4}$ when it is constant as shown in \cref{fig:Coherence_Revelation_Sensitivity_Scaling_Schematic}.

In \cref{fig:Solar halo dominating detection prospects} we show the parameter space where a Solar DM halo would dominate the detection prospects of a clock comparison search in terms of the DM mass and the Solar halo density  relative to the galactic density $\rho^\odot/\rho^\mathrm{gal}$. For the virial halo on the right we further indicate the density that a Solar halo would have if e.g. gravitational interactions \cite{Xu:2008ep,Peter:2009mm} would equilibrate its occupation numbers with the galactic halo.

It becomes clear that datasets extending over a decade as e.g. the ones reported in \cite{Filzinger:2023zrs}, may detect a Solar halo only contributing $\sim 1\%$ of the galactic halo first. This presents a natural extension of current searches for halos dominating the density \cite{Wilson:2025lhq}.
We expect potentially even larger enhancements in sensitivity for a DM halo around Earth, since all timescales for such a halo are shorter and the effect of recoherence therefore matters more over the period an experiment may be run. We leave a detailed investigation of this case to future work.


\begin{acknowledgments}
The authors thank A. Caputo and N. Rodd for 
comments on the manuscript and D. Budker and A. Sushkov for insightful discussions. GP is funded by the ISF, Minerva, NSF- BSF and the European Union (ERC, DM-Dawn, 101199868).

\end{acknowledgments}


\clearpage

\onecolumngrid

\begin{center}
\Large\bfseries Supplemental Material
\end{center}

\bigskip
\twocolumngrid
\section{\label{sec:Details derivation of cohernece time}Derivation of coherence regimes}

In this section, we provide a qualitative derivation of the different coherence regimes, relying on our definition of the generalized coherence time in \cref{eq:Coherence_fraction_discrete}. We start by converting the sum over modes into continuum energy integrals, weighting with the density of states. Proceeding,

\begin{equation}
    \frac{\tau_{\mathrm{coh}}}{T_{\mathrm{obs}}} = \frac{\int_{0}^{\infty}  \int_{0}^{\infty} dE_1 dE_2 \mathcal{K}(E_1,E_2)\, \sinc( (E_{1}-E_2) T_\mathrm{obs}) }{\int_{0}^{\infty}  \int_{0}^{\infty} dE_1 dE_2 \mathcal{K}(E_1,E_2) } \,, \label{eq:Coherence_fraction_DOS} 
\end{equation} where the kernel of integration is, $\mathcal{K}(E_1,E_2)=f(E_1) f(E_2) e^{ - \beta (E_1 + E_2)}$. We have assumed a Maxwell-Boltzmann distribution for the population of modes, with inverse temperature $\beta$, and defined a function $f(E)$, for the density of states. For simplicity, we will consider the case where the density of states is such that the temperature controls the typical spread in the energies, $\beta^{-1} \sim \Delta E$\,. We change the variables of integration to $S=E_1 + E_2$ and $D=E_1 - E_2$. In terms of these variables,  \\  \begin{equation}
    \frac{\tau_{\mathrm{coh}}}{T_{\mathrm{obs}}} = \frac{\int_{0}^{\infty} \int_{-S}^{S} dS dD \, \mathcal{K}\left(\left(  \frac{S+D}{2}\right),\left(  \frac{S-D}{2}\right)\right) \sinc( D T_\mathrm{obs}) }{\int_{0}^{\infty} \int_{-S}^{S} dS dD \, \mathcal{K}\left(\left(  \frac{S+D}{2}\right),\left(  \frac{S-D}{2}\right)\right)} \,. 
\end{equation}

The behavior of this integral depends crucially upon three energy scales, experimental resolution $ \sim 2 \pi / T_\mathrm{obs}$\,, temperature $ \sim 1/\beta$\,, and the smallest relevant energy spacing in the system $\sim \delta E$\,. We assume a high enough temperature to occupy multiple modes, $\beta \delta E \lesssim 1$. Therefore, we will have three different regimes based on the value of $2 \pi / T_\mathrm{obs}$. In each regime, we note that the $\sinc$ gives a $\mathcal{O}(1)$ contribution only if the argument is less than about $2 \pi$, $ D T_\mathrm{obs} \lesssim 2 \pi$, and the exponential contributes only when $\beta S \lesssim 1$. Keeping this in mind, we investigate the three regimes separately,

\subsection*{%
  \texorpdfstring
    {Coherence: $\delta E \lesssim \beta^{-1} \lesssim 2\pi/T_\mathrm{obs}$}
    {Coherence: delta E <= betapower(-1) <= 2pi/Tobs}%
}

Over the entire range of integration, the $\sinc$ gives a $\mathcal{O}(1)$ contribution, and thus the numerator is approximately equal to the denominator. This results in a linear scaling of the coherence time $\tau(T_\mathrm{obs}) \simeq T_{\mathrm{obs}}$\,.

\subsection*{%
  \texorpdfstring
    {Decoherence: $\delta E \lesssim 2\pi/T_\mathrm{obs} \lesssim \beta^{-1}$}
    {Decoherence: delta E <= 2pi/Tobs <= betapower-1}%
}
The $\sinc$ forces the integration range of $D$ to be constrained to, $(-2 \pi / T_\mathrm{obs},2 \pi / T_\mathrm{obs})$, as otherwise it leads to suppressed contributions. We may therefore approximate the integrals as  
\begin{equation}
    \frac{\tau_{\mathrm{coh}}}{T_{\mathrm{obs}}} \sim  \frac{\int_{0}^{1/\beta} \int_{-2 \pi / T_\mathrm{obs}}^{2 \pi / T_\mathrm{obs}} dS dD f\left(  \frac{S+D}{2}\right) f\left(  \frac{S-D}{2}\right) }{\int_{0}^{1/\beta} \int_{-1/\beta}^{1/\beta} dS dD f\left(  \frac{S+D}{2}\right) f\left(  \frac{S-D}{2}\right)} \,,
\end{equation}
where we dropped the $\mathcal{O}(1)$ contributions arising from the exponential and sinc function. We can further approximate $f\left(  \frac{S \pm D}{2}\right) \approx f\left( \frac{S}{2}\right)$, if the density of states is a smooth function. 
We observe that the $S$ integration factors cancel and the $D$ integration factors lead to a scaling $\tau(T_\mathrm{obs}) \sim 2\pi \beta$\,. Instead of growing linearly the coherence time is now constant $\sim 2\pi \beta$\, and we expect the transition near $2 \pi \beta / T_\mathrm{obs} \sim 1$\,. The height of this plateau is exactly the finite coherence time one expects from the conventional coherence time.
Let us now investigate what happens when the observation time is long enough to resolve the spacing $\delta E$.

\subsection*{%
  \texorpdfstring
    {Recoherence: $2\pi/T_\mathrm{obs} \lesssim \delta E \lesssim \beta^{-1}$}
    {Recoherence: 2pi/Tobs <~ delta E <~ betapower(-1)}%
}

Since the scale $2 \pi / T_\mathrm{obs}$\, is now the smallest in the problem, the sinc contributes only if $D$ is identically zero. Thus, we can approximate the scaling in this regime by replacing the sinc with a delta function, \begin{equation}
    \frac{\tau_{\mathrm{coh}}}{T_{\mathrm{obs}}} \sim  \frac{\int_{0}^{1/\beta} \int_{-2 \pi / T_\mathrm{obs}}^{2 \pi / T_\mathrm{obs}} dS dD f\left(  \frac{S+D}{2}\right) f\left(  \frac{S-D}{2}\right) \delta E \, \delta(D) }{\int_{0}^{1/\beta} \int_{-1/\beta}^{1/\beta} dS dD f\left(  \frac{S+D}{2}\right) f\left(  \frac{S-D}{2}\right)} \,, 
\end{equation}
where we introduced the energy spacing scale to restore the dimensions. Similarly as before, the $S$ integration factors cancel, and the $D$ integration factors lead to scaling of form, $\tau_\mathrm{coh}(T_{\mathrm{obs}}) \sim \beta \delta E \, T_{\mathrm{obs}}$\,. Therefore, we observe that the generalized coherence time again exhibits a linear rise, though with a much weaker slope. The coherent nature is arising because one now has the ability to probe individual modes, however, this naturally leads to the loss of collective amplitude enhancement. \\

In complex cases, such as the case of a sub-dominant Solar halo, where only part of the spectrum is quantized with typical spacing $\delta E$ while the rest of the spectrum is effectively continuous (spacing unresolvably small), the slope is further suppressed by $\approx (P_\mathrm{quant}/P_\mathrm{tot})^2$\,.
Here $P_\mathrm{tot/quant}$ denote the total signal power and the one in the quantized levels, respectively. The signal power is defined as $P_\mathrm{tot}=\sum_i |\psi_i|^2N_i$, while for  $P_\mathrm{quant}$ the sum is restricted to the quantized part of the spectrum.

\section{\label{sec:Supplementary information on the coherence examples}Details of the coherence examples}

\subsection{\label{sec:Coherence for free quantum gas in a non-degenerate 3D box}Free quantum gas in a non-degenerate 3D box}

Let us consider the same toy example as discussed in the main text, albeit with a cuboid box of unequal side lengths, $L_1,L_2,L_3$\,. We choose the lengths such that the relative ratios among their squares are irrational. For example, $L^2 = L^2_3 = \frac{L^2_2}{\sqrt{3}} = \frac{L^2_1}{\sqrt{2}}$, which lifts all the degeneracies that were present in the case of a cubic box. Therefore, we can estimate the total number of non-degenerate states as $N_\mathrm{nd} \sim (k_\mathrm{max}/k_{\mathrm{min}})^3 \sim \left( \sqrt{2m/\beta} / (2 \pi / L) \right)^3 = \left( \frac{2 \pi^2 \beta}{m L^2} \right)^{-3/2}$, where $k_\mathrm{max}=\sqrt{2m/\beta}$ is the mode with the largest momentum before exponential suppression due to the Boltzmann distribution sets in and $k_{\mathrm{min}}$ is the spacing on the momentum grid. The increased number of non-degenerate levels leads to a smaller typical energy spacing $\delta E \sim \beta^{-1} N_\mathrm{nd}^{-1}$\, and therefore delayed onset of recoherence $T_\mathrm{obs} \sim \beta \left( \frac{2 \pi^2 \beta}{m L^2} \right)^{-3/2}$ compared to the degenerate box $T_\mathrm{obs} \sim \beta \left( \frac{2 \pi^2 \beta}{m L^2} \right)^{-1}$. 

\begin{figure}[h!]
    \includegraphics[scale=0.40]{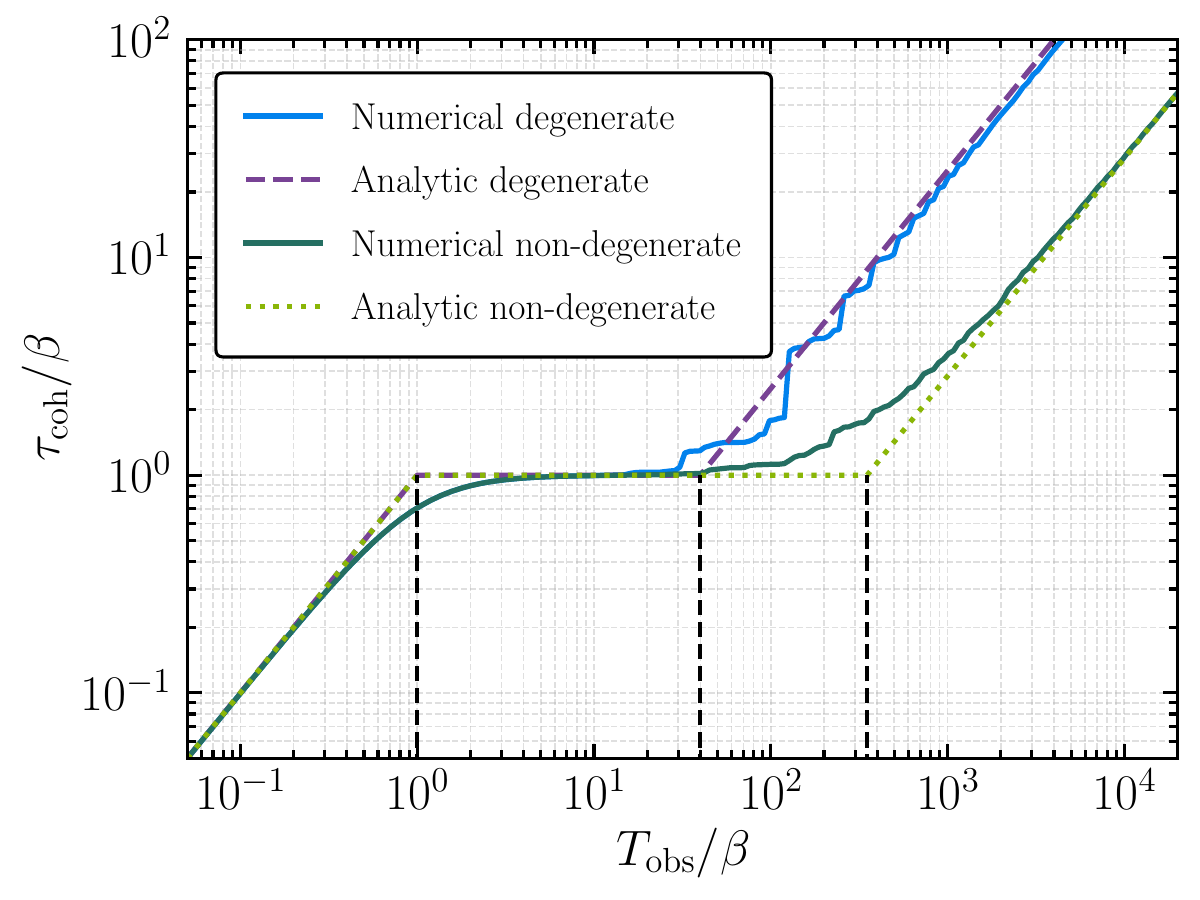}
     \caption{Evolution of the coherence time for a free quantum gas in a 3D box case with temperature $\beta^{-1}$ and size of the box $L$ chosen such that \,$\frac{2 \pi^2 \beta}{m L^2}=0.05$\,.  We compare the case of a degenerate box with a non-degenerate one, where the ratios between lengths is chosen as described in the text. In the non-degenerate case the onset of recoherence is delayed.}
    \label{fig:Free_3D_Box_Comparison_Coherence_Time_g_0.05}
\end{figure}

In \cref{fig:Free_3D_Box_Comparison_Coherence_Time_g_0.05} we show numerical evaluations of the generalized coherence time for both the degenerate and non-degenerate box and compare them to our analytic estimates. The delayed onset of recoherence in the non-degenerate case is clearly visible and we find good agreement between the estimates and the exact solutions away from the transition regions.

\subsection{\texorpdfstring{$1s$ Solar halo}{Sensitivity to the 1s Solar halo}}
\label{sec:sensitivity-1s-solar-halo}
\begin{figure}[h!]
    \includegraphics[scale=0.48]{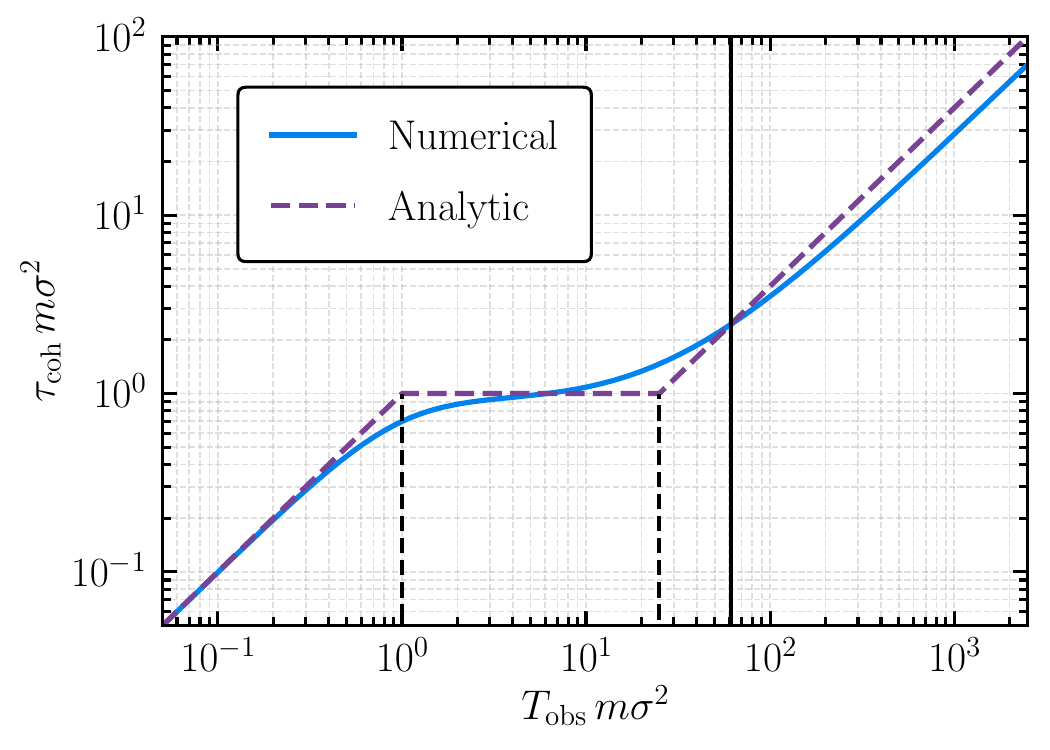}
     \caption{$1s$ Solar halo: Coherence time evolution for Solar halo ground state in the background of galactic DM for the point in parameter space marked with a cross in Fig.~\ref{fig:Solar halo dominating detection prospects}. The recoherence time is estimated as $T_\mathrm{obs}\sim (\rho_\mathrm{gal}/\rho^{\odot}_{\text{1s}})^2/(m\sigma^2)$. The vertical black dashed lines mark the de- and recoherence time, whereas the vertical thick black line correponds to $T_\text{obs}=1\,\text{year}$.}    \label{fig:Solar_mixed_ground_state_halo_coherence_time_plot}
\end{figure}

In \cref{fig:Solar_mixed_ground_state_halo_coherence_time_plot} we show the evolution of the coherence time for the case of a ground state halo around the Sun in the background of galactic DM. The mass and density ratios correspond to the point marked by a cross in Fig.~\ref{fig:Solar halo dominating detection prospects} on the left $m=2 \cdot 10^{-15}\,\mathrm{eV}$ and $\rho^\odot_{1s}/\rho^\mathrm{gal}=0.2$\,.

\subsection{\label{sec:Coherence for virialized Solar halo}Virialized Solar halo}

We provide here the analytic estimates for the virialized Solar halo. Accounting for the Earth's rotation around the Sun lifts up the degeneracy in magnetic quantum number. On Earth, the oscillations of $\psi_{nlm_l}$ are observed at frequencies $E'_{nlm_l}= - m \alpha^2 / 2 n^2+m_l\omega_\text{rot}$, where $\omega_\text{rot}=2\pi/\text{1 year}$\,. Therefore, every $n$ state now splits into $2n-1$ non-degenerate states. This leads to an increase in the typical number of non-degenerate states contributing significantly to the spectrum, $n_\text{typ} \sim n_{\mathrm{max}}\,(2 n_{\mathrm{max}} - 1)$. Accordingly, we can estimate the decoherence and recoherence time scales as
\begin{align}
    T_{\mathrm{obs,decoherence}}& = \frac{\pi}{2}\,\frac{n^2_{\mathrm{max}}}{m \alpha^2} \\ & =\frac{\pi}{2 m v^2} \\ T_{\mathrm{obs,recoherence}} & = \frac{\pi}{2}\,\frac{n^3_{\mathrm{max}} (2 n_{\mathrm{max}} - 1)}{m \alpha^2} \\ &=\frac{\pi}{2 m v^2} \left( \frac{2 r}{a_0} - \sqrt{\frac{r}{a_0}} \right) \,. 
\end{align}
We compare these scalings in \cref{fig:Solar halo dominating detection prospects} to a numerical evolution of the coherence time.

While the density profile for the $1s$ Solar halo is trivially exponential, the density profile for the virialized Solar halo can be well approximated by the  piecewise function
\begin{align}
    \rho(r) & = \rho(0) \exp(-2r/a_0)\,,\ r\leq a_0\\
    & = \rho(0) \, e^{-2} \left(\frac{a_0}{r}\right)^{3/2}\,,\ r>a_0 \,. 
\end{align} 
For a virial halo in equilibrium with the galactic halo one has $\rho(0)=(2\pi)^{3/2}/\pi \cdot (\alpha/\sigma)^{3}$\,, which is leads to the green line in \cref{fig:Solar halo dominating detection prospects}

\begin{figure}[th!]
    \includegraphics[scale=0.48]{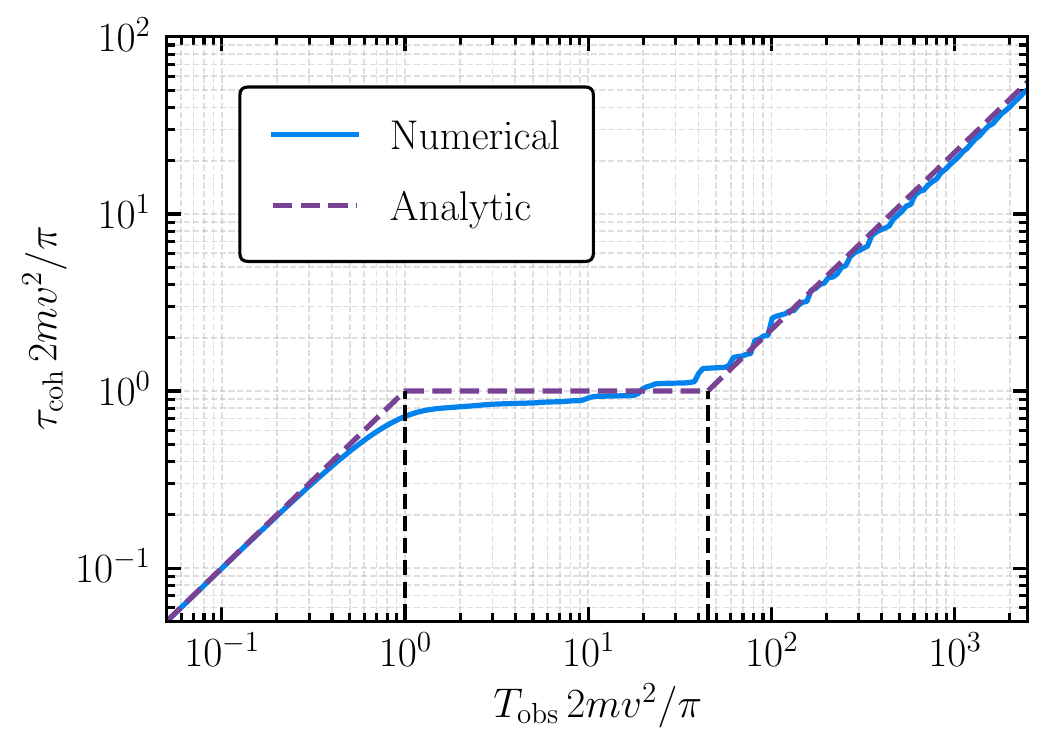}
     \caption{Evolution of the coherence time for the virialized Solar halo only with $r/a_0=25$. The spread of energies leads to a finite decoherence time $\sim 1/m v^2$, in agreement with the standard picture. The discrete nature of the bound states gives rise to recoherence at $T_\mathrm{obs} \sim \delta E^{-1} \sim (r/a_0)/m v^2$.
     }
    \label{fig:Virial_Solar_Halo_Coherence_Time_Nc_25}
\end{figure} 

\section{Derivation of SNR \label{app:SNR}}
In this section we follow \cite{Allen:1997ad,Thrane:2013oya,Breitbach:2018ddu,Romano:2016dpx} to derive the signal-to-noise ratio (SNR) for a stochastic signal. Consider a total data series $x(t)=s(t)+n(t)$, where $s$ denotes the signal and $n$ the noise. For example $x$ might be the measured frequency ratios in a clock comparison experiment. These data points are taken at discrete times $t_k=k \Delta t$ and in total we have $N=T/\Delta t$ of them, where $T$ is the total observation time. 

Given the Fourier coefficients $s(\omega)$ of the signal over all times that a model predicts, we may calculate the discrete Fourier coefficients for freqeuncy bins $\omega_k=k\Delta\omega$, where $\Delta\omega=2\pi/T$
\begin{align}
    s_k&=\sum_{m=0}^N \Delta t\,s(t_m)\exp(i\omega_k t_m)\\
    &=\sum_{m=0}^N \Delta t\int \frac{d\omega}{2\pi}s(\omega)\exp(i(\omega_k-\omega) t_m)\\
    &=\Delta t\int \frac{d\omega}{2\pi}s(\omega)\frac{\exp(i(\omega_k-\omega) T)-1}{\exp(i(\omega_k-\omega) \Delta t)-1}\\
    &\simeq2T\int \frac{d\omega}{2\pi}s(\omega)\exp\left(i(\omega_k-\omega) \frac{T}{2}\right)\mathrm{sinc}\left((\omega_k-\omega) \frac{T}{2}\right)\,.
\end{align}
We assumed in the last step that the spacing between measurements is small enough that all relevant signal frequencies are small compared to the experiments UV cut-off $(\omega_k-\omega) \Delta t\ll1$\,.

Given the power spectrum $P_\mathrm{sig}$ of the signal, defined by $\langle s(\omega_1)s^*(\omega_2)\rangle=2\pi\delta(\omega_1-\omega_2)P_\mathrm{sig}(\omega_1)$\,, we can calculate the correlation of the signal in two bins
\begin{widetext}
\begin{align}
    \langle s_k s_l^*\rangle&=4T^2\int\frac{d\omega_1\,d\omega_2}{(2\pi)^2}\langle s(\omega_1)s^*(\omega_2)\rangle\exp\left(i(\omega_k-\omega_l-\omega_1+\omega_2)\frac{T}{2}\right)\mathrm{sinc}\left((\omega_k-\omega_1)\frac{T}{2}\right)\mathrm{sinc}\left((\omega_l-\omega_2)\frac{T}{2}\right)\\
    &=4T^2\int\frac{d\omega}{2\pi}P_{sig}(\omega)\exp\left(i(\omega_k-\omega_l)\frac{T}{2}\right)\mathrm{sinc}\left((\omega_k-\omega)\frac{T}{2}\right)\mathrm{sinc}\left((\omega_l-\omega)\frac{T}{2}\right)\,.\label{eq: expectation signal}
\end{align}
\end{widetext}
One can carry out the same calculation for the noise. Assuming a white noise spectrum $P_\mathrm{ns}(\omega)=\mathrm{const.}$ one can carry out the integration in $\omega$ and finds
\begin{align}
    \langle n_k n_l^*\rangle&=4T\,P_{\text{ns}}\exp\left(i(\omega_k-\omega_l)\frac{T}{2}\right)\mathrm{sinc}\left((\omega_k-\omega_l)\frac{T}{2}\right)\\
    &=4T\,P_{\text{ns}}\delta_{kl}\,.
\end{align}
At this point one can define a test statistic
\begin{align}
    Y=\sum_{kl}x_lx^*_k\, Q^*_{kl} =\langle xx^*|Q\rangle\,,
\end{align}
where we introduced a scalar product in the second step as a short hand. $Q_{kl}$ is a filter function that is determined below to maximize the signal to noise ratio (SNR). The expectation value of the test statistic if there is only signal is
\begin{align}
    \langle Y\rangle_\mathrm{sig}=\langle \langle ss^*\rangle|Q\rangle\,.
\end{align}
The variance caused by the noise is
\begin{align}
    \langle Y^2\rangle_\mathrm{ns}-\langle Y\rangle^2_\mathrm{ns}&=2\sum_{klmn}\langle n_k n^*_n\rangle\langle n_m n^*_l\rangle Q_{kl}Q^*_{mn}\\
    &=2(4TP_\mathrm{ns})^2\langle Q|Q\rangle\,.
\end{align}
We can now introduce the signal to noise ratio
\begin{align}
    \mathrm{SNR}^2&=\frac{\langle Y\rangle_\mathrm{sig}^2}{\langle Y^2\rangle_\mathrm{ns}-\langle Y\rangle^2_\mathrm{ns}}\\
    &=\frac{1}{2(4TP_\mathrm{ns})^2}\frac{|\langle \langle ss^*\rangle|Q\rangle|^2}{\langle Q|Q\rangle}\,.
\end{align}
A geometric consideration gives that $Q\propto \langle ss^*\rangle$
maximizes the SNR. We have
\begin{align}
    \mathrm{SNR}^2
    &=\frac{1}{2(4TP_\mathrm{ns})^2}\langle \langle ss^*\rangle|\langle ss^*\rangle\rangle\,.
\end{align}
Plugging the result of \cref{eq: expectation signal} and carrying out the implicit sum in the scalar product one finds
\begin{align}
    \mathrm{SNR}^2
    &=\frac{T^2}{2}\int\frac{d\omega_1 d\omega_2}{(2\pi)^2}\mathrm{sinc}^2\left((\omega_1-\omega_2)\frac{T}{2}\right)\frac{P_\mathrm{sig}(\omega_1)P_\mathrm{sig}(\omega_2)}{P^2_\mathrm{ns}}\,.
\end{align}
As a sanity check we can consider the case of a smooth spectrum in which case we may approximate $\mathrm{sinc}^2\left((\omega_1-\omega_2)\frac{T}{2}\right)\simeq2\pi/T\, \delta(\omega_1-\omega_2)$ and find the well known result for broad spectra 
\begin{align}
    \mathrm{SNR}^2
    &=\frac{T}{2}\int\frac{d\omega}{2\pi}\frac{P^2_{sig}(\omega)}{P^2_{ns}}\,.
\end{align}
Comparing with our definition of the coherence time
\begin{align}
    \tau(T)&=2\int_0^{T} d\tau\frac{\Gamma^2(\tau)}{\Gamma^2(0)}\\
    &=\frac{2T}{\Gamma^2(0)}\int\frac{d\omega_1 d\omega_2}{(2\pi)^2}\mathrm{sinc}\left((\omega_1-\omega_2)T\right)P_{sig}(\omega_1)P_{sig}(\omega_2)
\end{align}
If we approximate $\mathrm{sinc}\left((\omega_1-\omega_2)T\right)\sim\mathrm{sinc}^2\left((\omega_1-\omega_2)\frac{T}{2}\right)/2$, since they both approximate a $\delta$-function for large $T$, we get
\begin{align}
    \mathrm{SNR}^2&\simeq\frac{T\tau(T)}{2}\frac{\Gamma(0)^2}{P_{ns}^2}\\
    &=\frac{T\tau(T)}{2}\left(\int\frac{d\omega}{2\pi}\frac{P_{sig}(\omega)}{P_{ns}}\right)^2\,.
\end{align}
This approximation holds in the sense that the reasoning from the derivation of the behavior of the coherence time still holds and the equation above therefore correctly captures the scaling of the SNR with the observation time.

\section{Comment on perturbations to the Solar basin \label{app:A comment on the Solar basin perturbations}}

In the main text, we approximate the bound states of the Solar halo as coherently oscillating, which is exact in the absence of interactions. When interactions or potential perturbations are present, the states acquire a finite width. For the recoherence analysis to remain valid, these widths must be sufficiently small, i.e. smaller than $(\mathrm{10 \,yr})^{-1}$\,. Below, we examine the relevant broadening effects and show that they are far below the inverse of any realistic observation time, and thus negligible.

First, following the mechanism discussed in Ref.~\cite{Budker:2023sex}, self-interactions of the axion field lead to the population of bound states on timescales of order gigayears, which will lead to widths $\sim (\mathrm{Gyr})^{-1} \ll (\mathrm{10 \,yr})^{-1}$\,, negligible for our study.

\begin{figure}[th!]
    \includegraphics[scale=0.48]{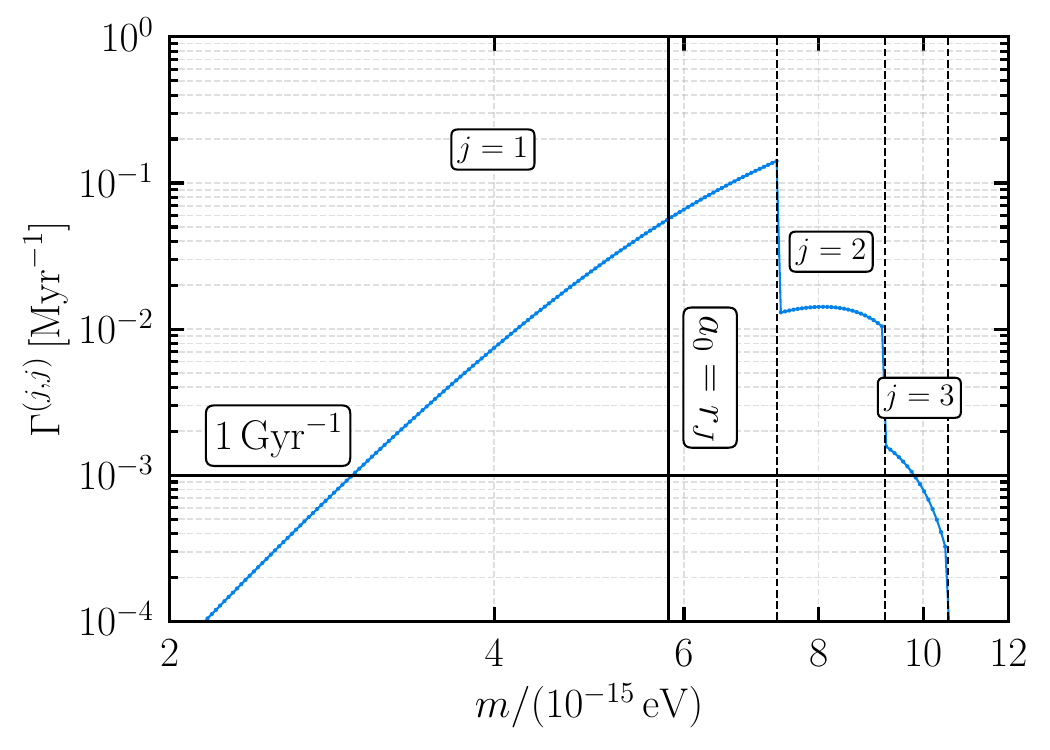}
     \caption{Leading multipole contribution to the ground state ionization rate induced by Jupiter's perturbation of the Solar basin. The vertical black dashed lines mark the regions corresponding to distinct leading multipoles. The vertical thick black line indicates the mass at which the Bohr radius equals the radius of Jupiter's trajectory, $a_0 = r_J$. The horizontal black line denotes the $1\,\mathrm{Gyr}$ lifetime. We observe that, in the mass range, $ 3 \cdot\,10^{-15}\,\mathrm{eV} \lesssim m \lesssim 10^{-14}\,\mathrm{eV}$, ionization due to Jupiter proceeds sufficiently rapidly to evaporate the ground state halo over the lifetime of the Sun.}
    \label{fig:Jupiter_Ionization_Numeric}
\end{figure}

Secondly, we consider perturbations to the Solar gravitational basin induced by Jupiter, the second most massive object in the Solar system, with mass ratio $M_J/M_\odot \sim 10^{-3}$\,. In the presence of Jupiter, the effective potential of the gravitational atom takes the form, \begin{equation}
    V(\vec{r},t) = -\frac{G_N M_\odot m}{r} - \frac{G_N M_J m}{\abs{\vec{r} - \vec{r}_J (t)}} = V(r) + \hat{V}(\vec{r},t) \,,
\end{equation} with $\vec{r}= r(\sin\theta \cos \phi , \sin\theta \sin\phi , \cos\theta)$ and $\vec{r}_J (t)$ denoting the Jupiter trajectory. We further assume that Jupiter follows a circular orbit in a plane perpendicular to the quantization axis, with constant angular frequency $\Omega_J$\,, such that its position vector is given by $\vec{r}_J= r_J (\cos(\Omega_J t),\sin(\Omega_J t),0)$\,, with $r_J \sim 5.2\,\mathrm{AU}$ being the average distance of the Jupiter from the Sun. We can evaluate the lifetime of the ground state due to the perturbation introduced by Jupiter, in accordance with Fermi's Golden rule,  
\begin{equation}
    \Gamma = \int \frac{d^3 \Vec{k}}{(2 \pi)^3} \lim_{t \to \infty} \frac{d}{dt} \abs{ i \int_{0}^{t} d t^\prime \,  \mathcal{M}(\Vec{k},t^\prime) \, e^{ - i \omega t^\prime} }^2  \,,
\end{equation} where, $\omega = k^2/2m + m \alpha^2/2$\,, is the total change in energy and the matrix element $\mathcal{M}$ is given as, \begin{equation}
    \mathcal{M}(\Vec{k},t^\prime) = \int d^3 \Vec{r} \, \psi^\ast_{100} (\Vec{r}) \, \hat{V}(\Vec{r},t^\prime) \,  \psi_{\Vec{k}} (\Vec{r}) \,.
\end{equation} We can further expand the perturbation in spherical harmonics and express the net ionization rate as a multipole expansion, \begin{equation}
    \Gamma = \sum_{l = 0}^{\infty} \sum_{m_l =-l}^{l} \Gamma^{(l,m_l)}  \,,
\end{equation} where the individual multipole contribution would contain a harmonic factor $\Gamma^{(l,m_l)} \supset e^{i m_l \Omega_J t^\prime}$ due to Jupiter's circular motion. Trivially the non-positive components, $m_l \leq 0$\,, lead to violation of energy conservation and thus do not contribute to the ionization. We can therefore re-express the sum as, \begin{equation}
    \Gamma = \sum_{m_l = j}^{\infty} \sum_{l=m_l}^{\infty} \Gamma^{(l,m_l)}  \,,
\end{equation} where, $j$ is the smallest positive integer which satisfies $j \geq m \alpha^2 / 2 \Omega_J$\,. In fact $j$ is the smallest multipole driving ionization and we have verified that it dominates the ionization rate, $\Gamma \approx \Gamma^{(j,j)}$\,. In \cref{fig:Jupiter_Ionization_Numeric} we show the results for this leading contribution $\Gamma^{(j,j)}$. 
This hierarchy between $\Gamma^{(j,j)}$ and the subleading correction $\Gamma^{(j+1,j+1)}$ gives rise to the observed discontinuities in the ionization rate whenever a given contribution $j$ becomes energetically forbidden.
From a physical standpoint, it is natural to expect that the perturbative effect induced by Jupiter is maximized when the Bohr radius of the atom is comparable to Jupiter’s orbital radius, $(m \alpha)^{-1} \sim r_J \sim 5.2\,\mathrm{AU}$. This expectation is in good agreement with the behavior observed in \cref{fig:Jupiter_Ionization_Numeric}. 

We conclude that the ionization-induced widths are smaller than $\mathrm{Myr}^{-1}$ and are therefore negligible on timescales relevant for any feasible measurement. However, the lifetime of a ground state Solar halo appears to be shorter than the age of the Solar System $\sim \mathrm{Gyr}$\, within the mass range $ 3 \cdot 10^{-15}\,\mathrm{eV} \lesssim m \lesssim 10^{-14}\,\mathrm{eV}$\,. Consequently, in this regime, the Solar halo cannot be attributed to specific initial conditions or dynamical processes operative at the time of Solar System formation. Instead, its existence necessitates dynamics occurring on significantly shorter timescales than $\mathrm{Myr}$. For larger masses, $m \gtrsim 10^{-14}\,\mathrm{eV}$\,, the thermalization effect induced by Jupiter is strongly suppressed and therefore does not play a role in the formation mechanism discussed in \cite{Budker:2023sex}.

Finally, we consider the effects of Solar flares. Typically, coronal mass ejections involve mass perturbations of order $\sim 10^{-18}\,M_\odot$, which are far too small to induce any appreciable widths in the Solar halo states.

\bibliography{apssamp}

\end{document}